\documentclass{revtex4}
\usepackage{amssymb}
\usepackage{amsmath}

\input{tcilatex}

\begin{document}

\title{One dimensional Newton's equation with variable mass}
\author{S. Habib Mazharimousavi}
\email{habib.mazhari@emu.edu.tr}
\author{M. Halilsoy}
\email{mustafa.halilsoy@emu.edu.tr}
\affiliation{Physics Department, Eastern Mediterranean University, G. Magusa north
Cyprus, Mersin 10 Turkey}

\begin{abstract}
We revisit Newton's equation of motion in one dimension when the moving
particle has a variable mass $m(x,t)$ depending both on position ($x$) and
time ($t$). Geometrically the mass function is identified with one of the
metric function in a $1+1-$dimensional spacetime. As a reflection of the
equivalence principle geodesics equation gives the Newton's law of motion
leaving the right hand side to be supplemented by the external forces. The
resulting equation involves the speed of light so that our equation of
motion addresses a wider scope than the customary classical mechanics. In
the limit of infinite light speed which amounts to instantaneous interaction
we recover the classical results.
\end{abstract}

\maketitle

\section{INTRODUCTION}

Numerous systems show that mass, which appears as a constant in the original
Newton equation need not be constant. A rolling snowball, launched rocket,
motion in a resistive medium etc., are just a few such examples that appear
in classical physics. Recently, there has been an increasing interest in
studying the position dependent mass (PDM) in classical systems at more
sophisticated level \cite{1,2,3,4,5,6}. Beside PDM some examples just cited
within the context of non-constant masses involve cases of time dependent
mass (TDM) as well. Herein, we wish to address both problems from a unified
and geometrical point of view.

The invariance group of classical mechanics, the Galilei group is no more an
invariance group in the presence of PDM. This reflects the fact that in the
frames in relative motion the accelerations are no more equal. As mass
depends on position such a situation creates different reaction forces to
different observers. Under this condition Galilei group in its standard form
doesn't provide a symmetry any more. Clearly PDM gives rise to velocity
dependent forces in the equation of motion. One such term familiar from the
mechanics text-books is the Reylegh's dissipation function \cite{7}. More
generally the mass can be considered as a function of both position and time
and the most general equation of motion can be derived from the
Euler-Lagrange (EL) formalism. In the absence of an external force EL method
serves well to derive the equation of motion. The physical meaning of
Hamiltonian, however, is no more considered as an energy, or at least not a
conserved energy and the Hamiltonian formalism is not well defined. In this
paper we derive, more generally the one-dimensional Newton's equation of
motion from a geometrical principle. Namely, we consider a $1+1-$dimensional
(i.e. one space, one time) line element of the form $ds^{2}=c^{2}dt^{2}-%
\frac{m\left( x,t\right) }{m_{0}}dx^{2},$ where $m\left( x,t\right) $ will
be interpreted as the mass function depending both on time ($t$) and
position ($x$) and $m_{0}=m\left( x=0,t=0\right) =const.$ Whenever $m\left(
x,t\right) $ depends only on position our 1+1-spacetime reduces
automatically to a flat space $ds^{2}=c^{2}dt^{2}-dX^{2},$ upon a scaling of
the coordinate $x.$ This yields the constant mass $m_{0}$ as an integration
constant. In general, for $m\left( x,t\right) $ we face a curved space whose
geodesics equation will be identical with the one-dimensional Newton's
equation of motion. For this purpose the affinely parametrized geodesics
equation with the proper time $\tau $ (i.e. $d\tau ^{2}=\frac{1}{c^{2}}%
ds^{2} $) must be transformed into the ordinary time ($t$).

The PDM geodesics equation takes the form $m\left( x\right) \frac{d^{2}x}{%
dt^{2}}+\frac{1}{2}\frac{dm}{dx}\left( \frac{dx}{dt}\right) ^{2}=0,$ which
is well known from other approaches. By this formalism we derive the
Newton's equation from a covariant, geometrical approach in which physical
mass emerges as a metric component of a curved space geometry. Our approach
is in conform with the equivalence principle of general relativity which
states the local equality of space curvature with the physical force
acceleration. No doubt in the limit of a constant mass, i.e. $m=m_{0}=$%
constant, we recover the well known limit of $m_{0}\frac{d^{2}x}{dt^{2}}=0,$
in the absence of an external force. In the presence of external forces of
any kind what one should do is just to modify our geodesics Lagrangian with
the supplementary terms.

\section{Time and position dependent mass}

Let's consider a one-dimensional classical system of mass $M$ which is
moving in positive $x-$direction with a speed of $v$. Also we assume that
there exists a time dependent force $F\left( t\right) =f_{0}\delta \left(
t\right) $ ($\delta \left( t\right) $ is the Dirac delta function) between
two parts of the original system with masses $M-m$ and $m.$ Here $f_{0}$ is
a constant to be found. The task of the force $F\left( t\right) $ is to
divide the original system into two parts with masses $m$ and $M-m$ at $t=0$
and the mass $m$ comes to rest for $t>0$ with respect to an observer at
rest. The reaction force i.e. $-F\left( t\right) $, acts on the other part
of the system to make its speed increasing. Here we only accept the Newton's
second law for constant mass particle which for mass $m$ reads, 
\begin{equation}
f_{0}\delta \left( t\right) =m\frac{dv}{dt}
\end{equation}%
and after integration becomes 
\begin{equation}
f_{0}\int_{-\epsilon }^{\epsilon }\delta \left( t\right) dt=m\int_{v}^{0}dv.
\end{equation}%
We note that the entire process takes place at $t=0$ and therefore $\epsilon 
$ is a positive constant. The latter equation gives 
\begin{equation}
f_{0}=-mv.
\end{equation}%
A similar equation is also applicable for the second part of mass $M-m$, 
\begin{equation}
-f_{0}\delta \left( t\right) =(M-m)\frac{dv}{dt}
\end{equation}%
which upon integration over the same time-interval yields%
\begin{equation}
-f_{0}\int_{-\epsilon }^{\epsilon }\delta \left( t\right)
dt=(M-m)\int_{v}^{v^{\prime }}dv.
\end{equation}%
Here $v^{\prime }$ is the speed of the mass $M-m$ after $t=0$ while $v$ is
its speed before $t=0.$ This equation gives%
\begin{equation}
-f_{0}=(M-m)\left( v^{\prime }-v\right)
\end{equation}%
which eventually yields%
\begin{equation}
\left( \frac{M}{M-m}\right) v=v^{\prime }.
\end{equation}%
Eq. (7) is nothing but 
\begin{equation}
Mv=\left( M-m\right) v^{\prime }
\end{equation}%
which by recalling that the mass $m$ is at rest shortly after $t=0$, this is
the linear momentum conservation. Having the velocities before and after $%
t=0 $, one can show that the total kinetic energy is not conserved i.e.,%
\begin{equation}
T_{f}-T_{i}=\left( \frac{m}{M-m}\right) \frac{1}{2}Mv^{2}\neq 0.
\end{equation}%
Now let's look at the problem from another perspective where due to a
position dependent internal force $F\left( x\right) =f_{0}\delta \left(
x\right) $, the same system (given above) is divided into two parts with
masses $M-m$ and $m$ at $x=0$. The function of this new force $F\left(
x\right) $ is to separate the mass $m$ from the original system and bring it
to rest wrt an observer at rest. The reaction force acts on the rest of the
mass i.e. $M-m$ to increase its speed to $v^{\prime }$ in the same
direction. The equation of motion for the mass $m$ is written as%
\begin{equation}
f_{0}\delta \left( x\right) =m\frac{dv}{dt}
\end{equation}%
and therefore%
\begin{equation}
f_{0}\int_{-\epsilon }^{\epsilon }\delta \left( x\right) dx=m\int_{v}^{0}vdv,
\end{equation}%
which admits%
\begin{equation}
f_{0}=-\frac{1}{2}mv^{2}.
\end{equation}%
A similar equation for $M-m$ yields%
\begin{equation}
-f_{0}\delta \left( x\right) =\left( M-m\right) \frac{dv}{dt}
\end{equation}%
or equivalently%
\begin{equation}
-f_{0}\int_{-\epsilon }^{\epsilon }\delta \left( x\right) dx=\left(
M-m\right) \int_{v}^{v^{\prime }}vdv.
\end{equation}%
This equation yields%
\begin{equation}
v^{\prime 2}=\left( \frac{M}{M-m}\right) v^{2}.
\end{equation}%
The latter equation is just the conservation of the kinetic energy while,
unlike the time-dependent mass system, the linear momentum is not conserved.
We note that having the mass $m$ at rest after switching on the internal
force $F\left( x\right) $ is just for simplicity and in the general approach
where $m$ may get a non-zero velocity, the results remain the same.

In conclusion, in the absence of any external force term, in the case of
time dependent mass classical system, the linear momentum is conserved but
kinetic energy not. In contrast, if the mass of the classical system is
position-dependent then the kinetic energy is conserved while the linear
momentum not.

What we have shown above clearly suggests that in the absence of any
external force term for a general time dependent mass particle one must use 
\begin{equation}
\frac{dp}{dt}=0
\end{equation}%
in which $p=M\left( t\right) v$, while for a general position-dependent mass
particle 
\begin{equation}
\frac{dp}{dt}\neq 0
\end{equation}%
with $p=M\left( x\right) v$ and instead one must use 
\begin{equation}
\frac{dT}{dt}=0
\end{equation}%
or equivalently 
\begin{equation}
M\left( x\right) \ddot{x}+\frac{1}{2}M^{\prime }\left( x\right) \dot{x}%
^{2}=0.
\end{equation}%
Here a prime "$^{\prime }$" denotes derivative wrt position $x$. Eq. (19)
can also be found by using the Hamiltonian method or Lagrangian formalism in
which the Lagrangian is given by 
\begin{equation}
L=\frac{1}{2}M\left( x\right) \dot{x}^{2}
\end{equation}%
and the Euler-Lagrange equation reads 
\begin{equation}
\frac{d}{dt}\frac{\partial L}{\partial \dot{x}}=\frac{\partial L}{\partial x}%
.
\end{equation}%
Our simple examples given above can be easily extended into the general case
with external forces. In such case if the mass is time dependent one should
start with 
\begin{equation}
\frac{dp}{dt}=F_{ext}
\end{equation}%
and for the particle with PDM the correct approach is introducing the
Lagrangian $L=\frac{1}{2}M\left( x\right) \dot{x}^{2}-V_{ext}\left( x\right) 
$ and then use the EL equation. We also add here that Eq. (22) is found if
the departed mass comes to rest. This means that if the mass $m$ does not
come to rest after the separation, we must consider $p$ the total linear
momentum of the system. This is the method which is used to find the
equation of a rocket.

Another remarkable point is that for a system with PDM usually we don't talk
about the departed mass, i.e., we only study the original system whose mass
is decreasing or increasing. This implicitly implies that the mass
difference (apart mass) is at rest and its linear momentum and kinetic
energy both are zero.

Finally we refer to Ref. \cite{6} where Eq. (3) (of \cite{6}) has been found
by considering the conservation of the linear momentum of a PDM particle in
absence of an external force. As we shown above the correct equation is Eq.
(19) and therefore by comparison, the proper form of the PDM function in
Ref. \cite{6} is given by 
\begin{equation}
m\left( x\right) =\frac{m_{0}}{1+\xi x^{2}},
\end{equation}%
in which $m_{0}$ has the meaning of a constant mass for $\xi =0.$

\section{A geometric model of PDM}

In this section we start with a $1+1-$dimensional curved spacetime in the
form of%
\begin{equation}
c^{2}d\tau ^{2}=ds^{2}=c^{2}dt^{2}-\frac{m\left( x,t\right) }{m_{0}}dx^{2}
\end{equation}%
or equivalently $d\tau ^{2}=dt^{2}-\frac{m\left( x,t\right) }{\alpha _{0}}%
dx^{2}$ where $\alpha _{0}=m_{0}c^{2}=const.$ Here $m\left( x,t\right) $ is
an arbitrary function of time and space while $m_{0}$ is a constant
representing $m_{0}=m\left( x=0=t\right) $ with identical dimension as $%
m\left( x,t\right) $. The geodesic equation of a free particle moving in
this spacetime is given by%
\begin{equation}
\frac{d^{2}x^{i}}{d\tau ^{2}}+\Gamma _{ab}^{i}\frac{dx^{a}}{d\tau }\frac{%
dx^{b}}{d\tau }=0
\end{equation}%
in which $\Gamma _{ab}^{i}$ is the Christoffel symbol, $\tau $ is the proper
time, $c$ is the speed of light and $i,a$ and $b$ run from $0$ to $1$. The
explicit equations of motions are then given by%
\begin{equation}
\frac{d^{2}x}{d\tau ^{2}}+\frac{m^{\prime }}{2m}\left( \frac{dx}{d\tau }%
\right) ^{2}+\frac{\dot{m}}{m}\left( \frac{dt}{d\tau }\right) \left( \frac{dx%
}{d\tau }\right) =0
\end{equation}%
and%
\begin{equation}
\frac{d^{2}t}{d\tau ^{2}}+\frac{\dot{m}}{2\alpha _{0}}\left( \frac{dx}{d\tau 
}\right) ^{2}=0.
\end{equation}%
Eliminating proper time, one finds a single equation 
\begin{equation}
m\frac{d^{2}x}{dt^{2}}+\dot{m}\frac{dx}{dt}+\left( \frac{m^{\prime }}{2}-%
\frac{m\dot{m}}{2\alpha _{0}}\frac{dx}{dt}\right) \left( \frac{dx}{dt}%
\right) ^{2}=0,
\end{equation}%
or equivalently 
\begin{equation}
\frac{d}{dt}\left( m\dot{x}\right) =\frac{1}{2}\dot{x}^{2}\left( m^{\prime }+%
\frac{m\dot{m}}{\alpha _{0}}\dot{x}\right) 
\end{equation}%
which follows directly from the variational principle $\delta \dint \sqrt{1-%
\frac{m\left( x,t\right) }{\alpha _{0}}\dot{x}^{2}}dt=0$. First of all let's
consider $m_{0}$ to be a constant so that%
\begin{equation}
m_{0}\frac{d^{2}x}{dt^{2}}=0
\end{equation}%
which is the Newton's second law for a free particle with mass $m.$ Then we
consider a time dependent mass which reads%
\begin{equation}
m\frac{d^{2}x}{dt^{2}}+\dot{m}\frac{dx}{dt}-\frac{m\dot{m}}{2\alpha _{0}}%
\left( \frac{dx}{dt}\right) ^{3}=0.
\end{equation}%
This equation in Newtonian limit (i.e. $c\rightarrow \infty $) becomes%
\begin{equation}
m\frac{d^{2}x}{dt^{2}}+\dot{m}\frac{dx}{dt}=0
\end{equation}%
or equivalently%
\begin{equation}
\frac{d\left( m\dot{x}\right) }{dt}=0
\end{equation}%
which is the correct Newton's second law for a free particle with time
dependent mass. In these two cases one also observes that the usual
definition of linear momentum i.e., $p=m\dot{x}$, implies that the linear
momentum is conserved. As a matter of fact (32) admits a first integral
given by 
\begin{equation}
\frac{1}{p^{2}}-\frac{1}{p_{0}^{2}}=\frac{1}{\alpha _{0}}\left( \frac{1}{m}-%
\frac{1}{m_{0}}\right) 
\end{equation}%
where $p_{0}$ and $m_{0}$ are constants of momentum and mass respectively.
This states openly that both for $m=m_{0}$ and $c\rightarrow \infty $ limits
we obtain $p=p_{0}$, i.e. conservation of momentum. Finally we consider the
case in which the function $m$ is only a function of position which admits%
\begin{equation}
m\frac{d^{2}x}{dt^{2}}+\frac{m^{\prime }}{2}\left( \frac{dx}{dt}\right)
^{2}=0.
\end{equation}%
This is the correct equation of motion for a free particle whose mass is
position dependent. We note that this equation can be written as%
\begin{equation}
\frac{d\left( m\dot{x}\right) }{dt}=\frac{m^{\prime }}{2}\left( \frac{dx}{dt}%
\right) ^{2}.
\end{equation}%
This clearly shows that the usual linear momentum is not conserved due to a
force $\frac{m^{\prime }}{2}\left( \frac{dx}{dt}\right) ^{2}$ which has
emerged from the PDM nature of the particle. In other words, the PDM changes
the geometry of spacetime in such a way that the particle experiences a new
geometrical force $F_{G}=$ $\frac{m^{\prime }}{2}\left( \frac{dx}{dt}\right)
^{2}.$ Furthermore, once the mass is only position dependent, the spacetime
is not any more curved and by a simple transformation 
\begin{equation}
\sqrt{\frac{m\left( x\right) }{m_{0}}}dx=dX
\end{equation}%
one gets%
\begin{equation}
ds^{2}=c^{2}dt^{2}-dX^{2}.
\end{equation}%
The geodesic equation of the free particle in this transformed spacetime
simply reads%
\begin{equation}
\frac{d^{2}X}{dt^{2}}=0,
\end{equation}%
or equivalently 
\begin{equation}
\frac{d}{dt}\left( \sqrt{\frac{m\left( x\right) }{m_{0}}}\frac{dx}{dt}%
\right) =0.
\end{equation}%
This in turn implies%
\begin{equation}
\frac{1}{2}m\left( x\right) \left( \frac{dx}{dt}\right) ^{2}=const.
\end{equation}%
which amounts to the conservation of kinetic energy. We note that 
\begin{equation}
X=\dint\nolimits^{x}\sqrt{\frac{m\left( y\right) }{m_{0}}}dy.
\end{equation}%
which for the case of the mass given in (23) one finds%
\begin{equation}
X=\frac{1}{\sqrt{\xi }}\sinh ^{-1}\left( \sqrt{\xi }x\right) .
\end{equation}%
One easily observes that for $\xi \rightarrow 0,$ one recovers, $%
ds^{2}=c^{2}dt^{2}-dx^{2},$ which yields the standard Newton's equation of
motion. Let's add that having external interaction would only lead to some
external forces to the right side which is a trivial extension of the model.
Finally we comment on the canonical formalism of our geometrical model.
Fixing the square-root Lagrangian with the correct dimension of energy we
have%
\begin{equation}
L\left( x,\dot{x}\right) =\sqrt{\alpha _{0}}\sqrt{\alpha _{0}-m\left(
x,t\right) \dot{x}^{2}}.
\end{equation}%
Canonical momentum is defined by $p=\frac{\partial L}{\partial \dot{x}}$ and
the usual Hamiltonian $H=p\dot{x}-L$ gives 
\begin{equation}
H\left( x,p\right) =\alpha _{0}\sqrt{1-\frac{p^{2}}{m\alpha _{0}}}.
\end{equation}%
It can be checked that $H\left( x,p\right) $ and $L\left( x,\dot{x}\right) $
satisfy the constraint 
\begin{equation}
H\left( x,p\right) L\left( x,\dot{x}\right) =1
\end{equation}%
and the usual Hamiltonian equation, $-\dot{p}=\frac{\partial H}{\partial x},$
doesn't give the correct equation of motion. This means that in such a
formalism the standard Hamilton equation, without further modifications,
fail to work while EL equation always works.

\section{Conclusion}

Einstein's general relativity is a geometric description of the universe and
is known to extend Newton's gravitational theory. Existence of the
equivalence principle at the very heart of general relativity transcends
physical theories from gravitation to the realm of all forces of nature.
From this token the recently fashionable PDM / TDM in various physics
theories is embedded into the geometric description of flat / curved
spacetime. In such a model mass can naturally be identified with the metric
function as the coefficient of spacetime element and as a result our model
lies in between special relativity and classical mechanics. Geodesic
equation is shown to yield the Newton's equation of motion with variable
mass in $1-$space dimension correctly. When the mass is variable both in $x$
and $t$ the resulting geodesic equation (or Newton's equation of motion)
becomes rather involved since it gives velocity dependent forces. These
forces are not put artificially but arise naturally as a result of our
geometric description. The equation obtained from the variational principle
are neither Galilean nor Lorentz invariant. Violation of these symmetries is
all due to the PDM and TDM appearing in the Lagrangian. It has been shown
that PDM gives conserved kinetic energy. As expected, taking mass as $%
m\left( x,t\right) $ leaves us with no conserved quantities and a highly
non-linear equation of motion. Undoubtedly, extending, this $1-$dimensional
toy model to $3-$dimensional Newton's equations with the effect of rotation
taken into account requires much more effort.

\end{document}